\documentclass[
    aps,
    prl,
    reprint,
    superscriptaddress,
    nofootinbib
]{revtex4-2}

\usepackage[utf8]{inputenc}
\usepackage[T1]{fontenc}
\usepackage{amsmath,amssymb,amsfonts,bm}
\usepackage{graphicx}
\usepackage{booktabs}
\usepackage{hyperref}
\usepackage{xcolor}
\usepackage{verbatim}
\usepackage[normalem]{ulem}

\hypersetup{
    colorlinks=true,
    linkcolor=blue,
    citecolor=blue,
    urlcolor=blue
}

\newcommand{\dd}{\mathrm{d}}

\newcommand{\varphidef}{\varphi}

\newcommand{\RF}{R_{\rm F}}
\newcommand{\RB}{R^{99}_{\rm B}}

\begin{document}

\title{Bosonic Dark Matter Cores Can Screen Scalar-Tensor Gravity}

\author{Jorge Castelo Mourelle}
\email{jorge.castelo@correo.nucleares.unam.mx}
\affiliation{
Instituto de Ciencias Nucleares,
Universidad Nacional Aut\'onoma de M\'exico,
Circuito Exterior C.U.,
A.P. 70-543,
Ciudad de M\'exico 04510,
M\'exico
}

\author{Saeed Fakhry}
\email{saeed.fakhry@uv.es}
\affiliation{
Departamento de Astronomía y Astrofísica,
Universitat de València,
Avinguda Vicent Andrés Estellés 19,
46100 Burjassot (València),
Spain
}

\author{Néstor Ortiz}
\email{nestor.ortiz@nucleares.unam.mx}
\affiliation{
Instituto de Ciencias Nucleares,
Universidad Nacional Aut\'onoma de M\'exico,
Circuito Exterior C.U.,
A.P. 70-543,
Ciudad de M\'exico 04510,
M\'exico
}

\date{\today}

\begin{abstract}
We examine the observational viability of scalarized fermion-boson stars (FBSs) in massless Damour-Esposito-Far\`ese gravity, using the MPA1 nuclear equation of state and a self-interacting complex scalar field playing the role of an ultralight bosonic dark matter model. Constraining stellar models within a multi-messenger-motivated mass-radius window, we obtain three main results: First, bosonic dark matter acts as a catalyst for spontaneous scalarization. Second, energetically favored configurations are scalarized FBSs featuring bosonic dark matter cores that reconcile strongly coupled regimes with radius bounds. Third, dark matter cores tend to suppress scalar charges, which mimics the effect of a massive gravitational scalar field and could contribute to evade observational constraints. Additionally, the exclusive prediction of scalarized FBSs at high masses suggests a way to test this theoretical scenario.
\end{abstract}

\maketitle

\textit{Introduction.—}
Investigating the nature of dark matter or possible deviations from General Relativity in the strong-field regime represents a cutting-edge endeavor in modern physics. Among alternative theories of gravity, the Damour-Esposito-Far\`ese (DEF) scalar-tensor model provides a compelling framework where scalar couplings induce the non-perturbative phenomenon of \emph{spontaneous scalarization} in compact objects~\cite{Damour:1992we,Damour:1993hw,Damour:1996ke,Harada:1997mr,Salgado:1998sg,Doneva:2022ewd}.
Spontaneous scalarization in pure neutron stars (NSs) induces asymptotic scalar charges and predicts significant changes in mass-radius relations, compactness, as well as dipolar scalar radiation, whose observational absence has placed stringent bounds on the coupling parameter $\beta \gtrsim -4.5$ from binary pulsar timing measurements~\cite{Bertotti:2003rm,Freire:2012mg,Antoniadis:2013pzd,Shao:2017gwu,Niu:2021nic,Tuna:2022qqr}.
However, most studies evaluating constraints on modified gravity assume distributions of purely baryonic matter, overlooking possible gravitational interactions with bosonic dark matter,\footnote{Notably, recent works have studied spontaneous scalarization of NSs in the presence of fermionic dark matter~\cite{Rahimi:2024qjl,Tanaka:2026wxs}.} for instance ultralight axion-like particles or scalar dark matter candidates that could accumulate in stellar environments, forming composite fermion-boson stars (FBSs)~\cite{Henriques:1989ar,Henriques:1989ez,Jetzer:1990xa,Pisano:1995yk,deSousa:1995ye,Valdez-Alvarado:2012rct,DiGiovanni:2020frc,Nyhan:2022pda,Mourelle:2024qgo,Konstantinou:2024ynd,Valdez-Alvarado:2024eze,Cipriani:2025wem,Zhou:2025kku}.

In this Letter, we explore the observational viability of scalarized FBSs in massless DEF gravity. We report the construction of such hybrid compact star models in the Companion Paper~\cite{CMjOnFs2026b} in a simplified scenario. Here, to enhance physical realism, we describe nuclear matter using the tabulated MPA1 equation of state~\cite{Muther:1987xaa}, and we account for self-interactions within the bosonic field sector. We constrain the total mass and effective radius within an observational window derived from combined NICER measurements~\cite{miller2019psr,Legred:2021hdx} and LIGO-Virgo gravitational-wave events~\cite{abbott2017gw170817,LIGOScientific:2020aai}, and we determine the energetically favored FBSs within this window.

We uncover three key results: First, the inclusion of a self-interacting bosonic dark matter component lowers the threshold for spontaneous scalarization, enabling scalar field activation at smaller coupling magnitudes $\vert{}\beta_{\text{crit}}\vert{}$ than in pure NSs across the observational range. Second, energetically favored FBSs are located within the observational window; they are scalarized objects with \textit{bosonic dark matter cores} confined within fermionic fluids, and they are compact enough to fit in the observational window even in the strong-coupling regime. Third, these bosonic cores alter the source trace of the scalar field equation of motion, forcing the gravitational scalar profile $\varphi(r)$ to decay more rapidly at large distances than in pure scalarized NSs of equivalent total mass. Such a \textit{bosonic screening} tends to suppress the scalar charge $Q_\varphi$, reducing the effects of massless DEF gravity. This reduction partially mimics the effect of a massive gravitational scalar  field~\cite{PhysRevD.92.124016,Ramazanoglu:2016kul,Yazadjiev:2016pcb,Doneva:2016xmf,Sperhake:2017itk,Staykov:2018hhc,Rosca-Mead:2019seq,Rosca-Mead:2020bzt,Huang:2021tpu,Kuan:2023hrh,Asakawa:2023obq,Degollado:2024oyo,Lam:2024wpq}. Furthermore, at high masses within the observational window, massless DEF gravity predicts energetically favored scalarized FBSs, but no scalarized NSs, suggesting a pathway for probing the massless DEF theory and scalarized FBSs with bosonic dark matter cores.

\textit{Theoretical Setup.—}
We consider the massless DEF scalar-tensor theory of gravity with fermionic and bosonic matter sources interacting through gravity. In the Einstein frame, in units where the speed of light $c=1=\hbar$, the action is
\begin{align}
S = \frac{1}{16\pi G}\int d^4x \sqrt{-\tilde{g}} \left[ \tilde{R} - 2\tilde{\nabla}_\mu\varphi \tilde{\nabla}^\mu\varphi \right] \nonumber \\
+ S_{\rm F}[g_{\mu\nu}, \Theta_{\rm F}] + S_{\rm B}[g_{\mu\nu}, \Phi],
\end{align}
where $\tilde{R}$ is the Ricci scalar associated with the Einstein-frame metric $\tilde{g}_{\mu\nu}$, and $\varphi$ is a real-valued gravitational scalar field. Both matter sectors couple universally to the Jordan-frame metric $g_{\mu\nu} = a^2(\varphi)\tilde{g}_{\mu\nu}$, with the coupling function $a(\varphi) = \exp(\frac{1}{2}\beta\varphi^2)$~\cite{Damour:1993hw,Damour:1996ke}.

The fermionic sector represents cold, high-density nuclear matter described by a perfect fluid and the microphysical tabulated MPA1 equation of state~\cite{Muther:1987xaa}. The bosonic dark matter sector is modeled by a massive, complex scalar field $\Phi(t,r) = \phi(r)e^{-i\omega t}$ governed by the action
\begin{equation}
S_{\rm B} = -\int d^4x \sqrt{-g} \left[ \nabla_\mu\Phi^* \nabla^\mu\Phi + V(|\Phi|^2) \right],
\end{equation}
with potential $V(|\Phi|^2) = \mu^2|\Phi|^2 + \frac{\lambda}{2}|\Phi|^4$.
We consider a fixed ultralight boson mass $m_{\rm B}=1.34\times10^{-10}\,{\rm eV}$, which corresponds to $\mu^2=1.0$~\cite{CMjOnFs2026b}.
In turn, the dimensionless self-interaction parameter is defined by $\Lambda_4\equiv\lambda M_{\rm Pl}^2/m_{\rm B}^2$, with $M_{\rm Pl}=G^{-1/2}$. We fix $\Lambda_4 = 50$ (corresponding to $\lambda=\simeq6.0\times10^{-75}$), which places the bosonic sector within the strong self-interaction regime~\cite{Colpi:1986ye,Schunck:2003kk,Liebling:2012fv} and yields solar-mass bosonic components~\cite{DiGiovanni:2020frc,Valdez-Alvarado:2024eze}.

Variations of the action yield the massless DEF-Klein-Gordon system,
\begin{align}
\tilde{G}_{\mu\nu} = 8\pi G \tilde{T}_{\mu\nu} &+ 2\tilde{\nabla}_\mu\varphi\tilde{\nabla}_\nu\varphi - \tilde{g}_{\mu\nu}\tilde{\nabla}_\sigma\varphi\tilde{\nabla}^\sigma\varphi, \\
\tilde{\Box}\varphi &= -4\pi G \alpha(\varphi)\tilde{T},\\
\tilde\nabla_{\mu} \left[ a^{2}(\varphidef)\tilde\nabla^{\mu}\Phi \right] &=
a^{4}(\varphidef) \frac{\dd V}{\dd |\Phi|^{2}} \Phi,
\end{align}
where $\alpha(\varphi) \equiv d\ln a(\varphi)/d\varphi = \beta\varphi$, and $\tilde{T}$ denotes the trace of the \emph{total} energy-momentum tensor. We construct spherical equilibrium FBS models in massless DEF gravity by numerically solving the static limit of the system above, together with the hydrostatic equations. We compute multi-parametric families of solutions by scanning the central fermionic rest-mass density $\rho_{0c}$ and the central bosonic field amplitude $\phi_c$. All details can be found in the Companion Paper~\cite{CMjOnFs2026b}.

\textit{Observational Window.—}
To evaluate the physical viability of mixed FBSs in massless DEF gravity, we define an observational window in the mass-radius plane. This window is bounded by $1.4 M_\odot \le M_{\text{ADM}} \le 2.4 M_\odot$ and $10\text{ km} \le R \le 13\text{ km}$. As displayed in the End Matter (Fig.~\ref{fig:observational_bounds}), this window overlaps with combined multi-messenger constraints from NICER mass-radius estimates for pulsars PSR~J0030+0451
\cite{miller2019psr,Kini:2026rjx} and PSR~J0740+6620
\cite{Salmi:2024aum}, and LIGO-Virgo gravitational-wave events GW170817 and GW190425~\cite{abbott2017gw170817,LIGOScientific:2020aai}. Recent analyses yield $R=12.68^{+1.31}_{-1.04}\,\mathrm{km}$ for PSR~J0030+0451 at $M=1.43^{+0.20}_{-0.17}\,M_\odot$~\cite{Kini:2026rjx}, and $R=12.49^{+1.28}_{-0.88}\,\mathrm{km}$ for PSR~J0740+6620 at $M=2.073\pm0.069\,M_\odot$~\cite{Salmi:2024aum}. Complementary constraints from GW170817 favor radii around $12\,\mathrm{km}$ \cite{LIGOScientific:2018cki}, while optical observations of PSR~J0952$-$0607 yield $M=2.35\pm0.11\,M_\odot$, motivating the upper mass limit \cite{Romani:2025ytn}. The adopted window is a broad selection criterion rather than a joint observational credible region; in particular, it does not exhaust the radii allowed by NICER. We also note that the NICER and LIGO-Virgo inferences assume General Relativity, and they could shift in DEF gravity.

\begin{figure}[t]
\centering
\includegraphics[width=\linewidth]{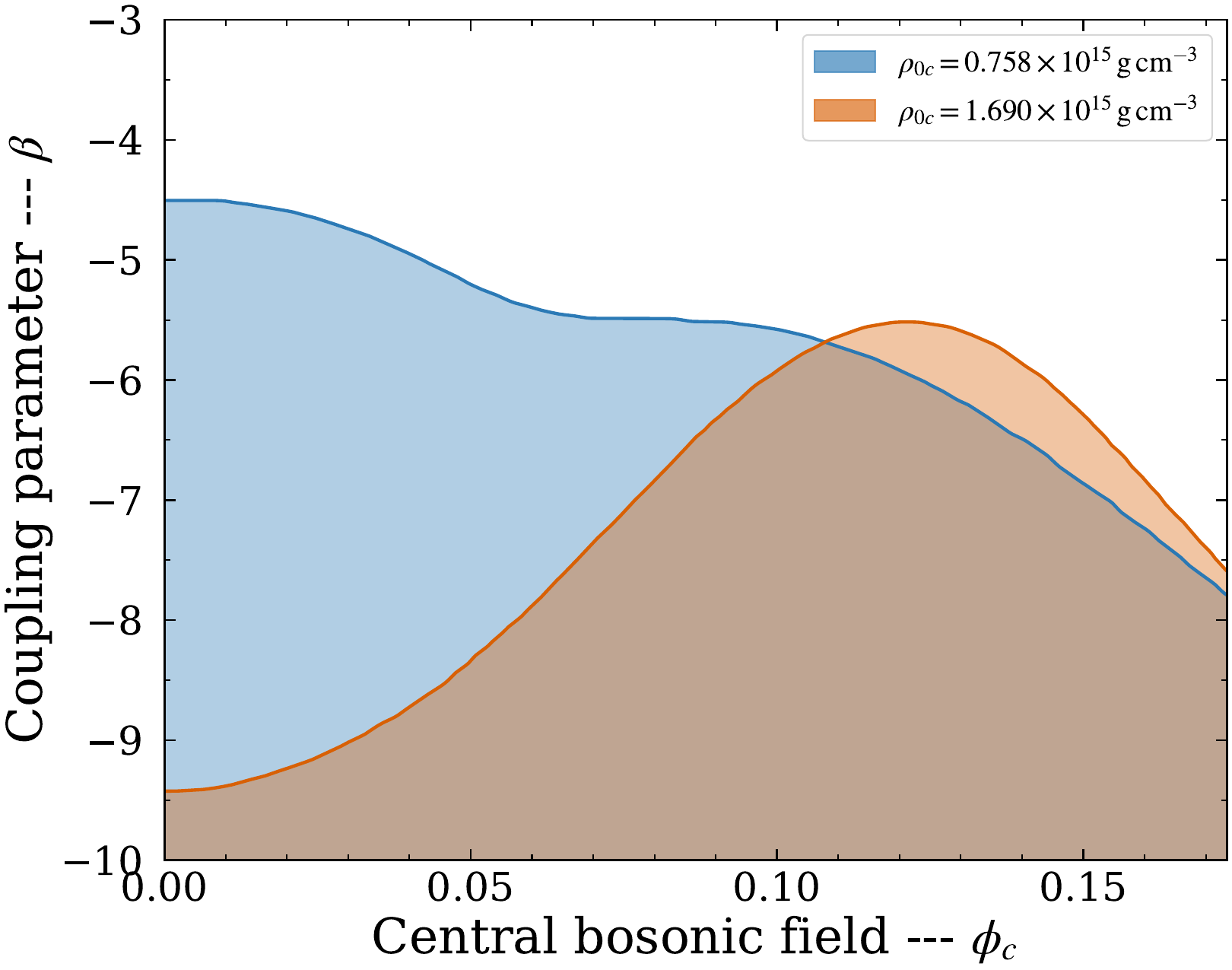}
\caption{Parameter space for scalarization of fermion-boson stars in this work. The blue and orange regions correspond to scalarized configurations at the central rest-mass densities that bound the observational window. The overlap of shaded regions indicates the existence of scalarized solutions \emph{within} the window, and the intersection boundary marks the critical coupling parameter $\beta_{\rm crit}(\phi_c)$ for the onset of spontaneous scalarization. As the central bosonic field $\phi_c$ increases, the barrier for scalarization, $|\beta_{\rm crit}|$, decreases relative to the pure neutron star case ($\phi_c=0$).}
\label{fig:beta_crit}
\end{figure}

\textit{Onset of Scalarization.—}
The presence of the bosonic sector alters the instability threshold for spontaneous scalarization. Figure~\ref{fig:beta_crit} shows the parameter space of coupling $\beta$ and central bosonic field amplitude $\phi_c$. Shaded regions indicate scalarized solutions for fixed central rest-mass densities $\rho_{0c} = 0.758 \times 10^{15}\text{ g cm}^{-3}$ (blue) and $\rho_{0c} = 1.690 \times 10^{15}\text{ g cm}^{-3}$ (orange), which correspond to the observational window bounds, as well as $\phi_c \in [0,0.1733]$. Details can be found in the End Matter. The overlap of shaded regions in Fig.~\ref{fig:beta_crit} corresponds to scalarized solutions \emph{within} the observational window, and the intersection boundary defines the critical coupling $\beta_{\text{crit}}(\phi_c)$ for the onset of scalarization. We observe that $\beta_{\text{crit}}$ increases from $\simeq -9.4$ in the pure NS case ($\phi_c = 0$), reaching $\beta_{\text{crit}} \approx -5.7$ around $\phi_c \simeq 0.11$. This shows that the bosonic field can act as a catalyst for spontaneous scalarization, lowering the critical barrier $|\beta_{\text{crit}}|$ required for scalar field activation in compact stars. Based on this result, in the following we analyze the coupling range $\beta \in [-10,-6]$.

\begin{figure}[t!]
    \centering
     \includegraphics[width=\columnwidth]{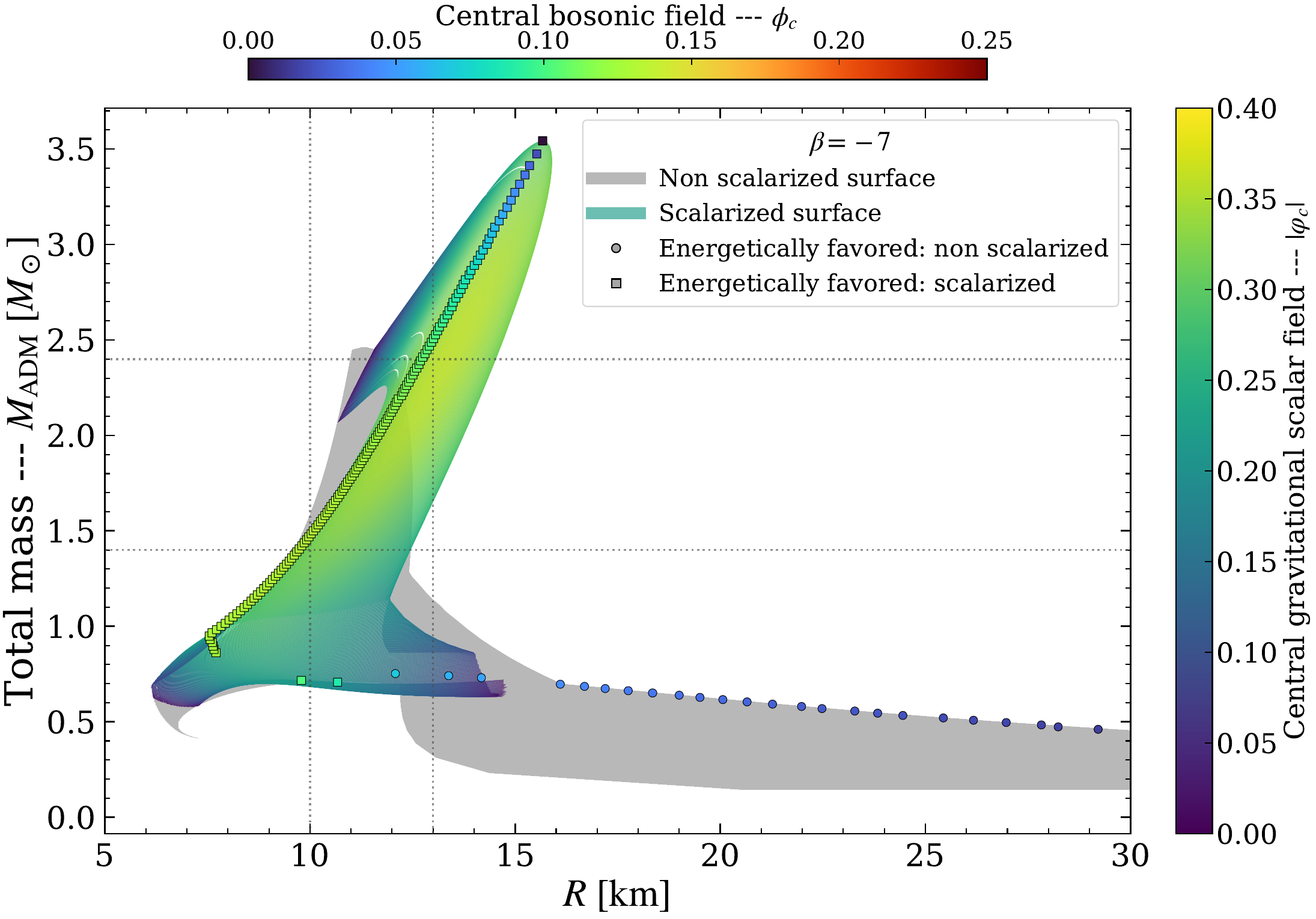}
       
    \caption{Mass-radius diagram for fermion-boson stars constructed with $\beta=-7$. The gray background represents non-scalarized solutions, whereas the colored surface corresponds to scalarized models, with the color indicating the absolute value of the central gravitational scalar field, $|\varphi_c|$. Circles (squares) identify energetically favored non-scalarized (scalarized) configurations, obtained by minimizing the total gravitational mass $M_{\rm ADM}$ at fixed total rest mass; their color denotes the central bosonic field $\phi_c$. Vertical and horizontal lines delimit the reference observational window, $10\leq R/{\rm km}\leq13$ and $1.4\leq M_{\rm ADM}/M_\odot\leq2.4$.}
    \label{fig:M-R-beta-7}
\end{figure}

\textit{Mass-radius relations.—}
As a representative scenario, we consider $\beta = -7$. In this case, Fig.~\ref{fig:M-R-beta-7} displays the two-dimensional projected mass-radius equilibrium surfaces. The surface of non-scalarized solutions (grey region) is significantly extended by spontaneous scalarization (colored region), which elevates the maximum supported total mass from $\sim 2.5 M_\odot$ up to $\sim 3.5 M_\odot$. The dependence of mass-radius relations on the coupling parameter $\beta$ is shown in the End Matter (Fig.~\ref{fig:M-R-different_beta}).

\textit{Energetically favored configurations.—}
Energetically favored configurations correspond to states that minimize the total gravitational mass, $M_{\text{ADM}}$, for a given rest mass, $M_0$, or equivalently, maximize the binding energy $E_{\text{bind}} = M_{\text{ADM}} - M_0 < 0$. See the End Matter for details. A slightly perturbed configuration---unstable to scalar field perturbations yet resilient to prompt gravitational collapse---should dynamically evolve toward a stable state of minimum $M_{\rm ADM}$.

\begin{figure}[t!]
    \centering
     \includegraphics[width=\columnwidth]{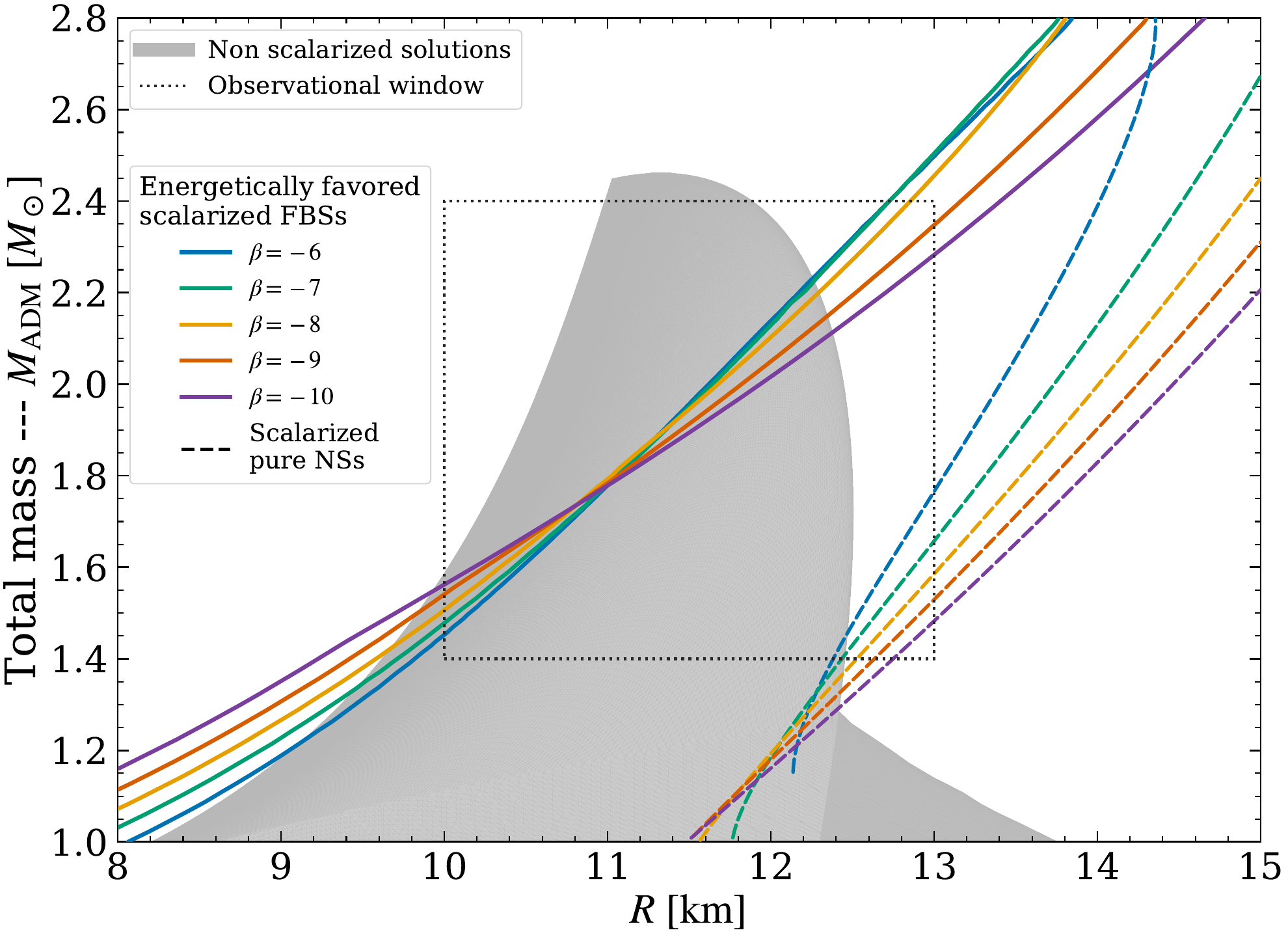}
       
    \caption{Mass-radius diagram comparing energetically favored scalarized fermion-boson stars with scalarized pure neutron stars, for $\beta\in\{-6,-7,-8,-9,-10\}$. Colored solid curves represent energetically favored mixed configurations, while dashed curves show the corresponding scalarized pure neutron-star sequences. The gray region contains non-scalarized solutions, and the dotted rectangle marks the reference observational window $10\leq R/{\rm km}\leq13$ and $1.4\leq M_{\rm ADM}/M_\odot\leq2.4$. Pure neutron-star sequences are displayed only up to their first crossing of either $R=15\,{\rm km}$ or $M_{\rm ADM}=2.8\,M_\odot$, thereby omitting ``returning'' branches. }
    \label{fig:window-several-betas}
\end{figure}

Across our observational window, scalarized FBSs are systematically preferred energetically over non-scalarized counterparts, establishing their physical relevance as ground-state candidates. As shown in Fig.~\ref{fig:M-R-beta-7}, the energetically preferred FBS equilibrium states trace a continuous sequence that passes through the observational window.  The parametric robustness of this result is shown in the End Matter (Fig.~\ref{fig:M-R-different_beta}), and summarized in Fig.~\ref{fig:window-several-betas}, where, for every $\beta \in \{-6, -7, -8, -9, -10\}$, the energetically favored scalarized FBS sequence consistently passes through the observational window. In comparison, pure scalarized NSs (dashed lines) of equivalent total mass are predicted to have larger radii.

Energetic comparisons at fixed rest mass provide useful indications of which configurations may be favored, but they do not establish dynamical stability. Ultimately, fully nonlinear simulations will be required to determine the fate of unstable and energetically disfavored configurations, including migration between branches, scalarization and descalarization, collapse, or partial bosonic dispersion through gravitational cooling~\cite{Seidel:1993zk}.

\textit{Compatibility with Radius Bounds.—}
A crucial insight emerges when inspecting the mass-radius behavior across different coupling strengths $\beta \in [-10, -6]$ in Fig.~\ref{fig:window-several-betas}. For strongly negative couplings, pure scalarized NS sequences (dashed lines) shift to radii progressively outside the observational window. In the absence of a bosonic component, such large coupling values, $\beta \lesssim -10$, would be ruled out by radius constraints alone. However, the gravitational binding of the bosonic component reduces the effective radius of energetically favored FBSs. Thus, the presence of bosonic dark matter would enable coupling parameter values that would otherwise be excluded by NS radius bounds. However, these results address radius constraints only. At fixed $M_{\rm ADM}$, FBSs exhibit a $22$-$29\%$ reduction in $\alpha_A\equiv|Q_\varphi|/M_{\rm ADM}$ relative to pure NSs, but whether this alleviates the tension with binary-pulsar constraints ($\beta\gtrsim-4.5$) requires a dedicated binary-dynamics analysis; see Table~\ref{tab:equal_adm_three_masses_beta6_beta7_beta8}.

\begin{figure}[t!]
    \centering
     \includegraphics[width=\columnwidth]{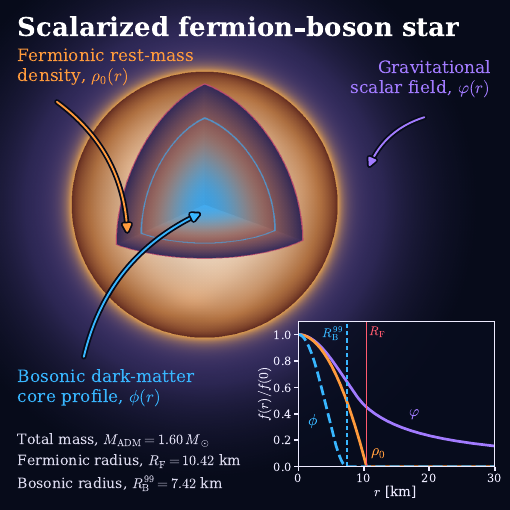}
    \caption{Radial profiles of a spherical, scalarized fermion-boson star in Damour-Esposito-Far\`ese scalar-tensor gravity, featuring a bosonic dark matter core. This solution is constructed with the nuclear equation of state MPA1, coupling parameter $\beta=-7$, boson mass $m_{\rm B}=1.34\times10^{-10}$ eV, and self-interacting parameter $\Lambda_4=50$. The bosonic (fermionic) radius is $\RB=7.42$ km ($\RF=10.42$ km), the total mass is $M_{\rm ADM}=1.60M_\odot$, and the scalar charge is $Q_\varphi=0.9899M_\odot$. 
    }
    \label{fig:3D_plot}
\end{figure}

\begin{figure}[t!]
    \centering
     \includegraphics[width=\columnwidth]{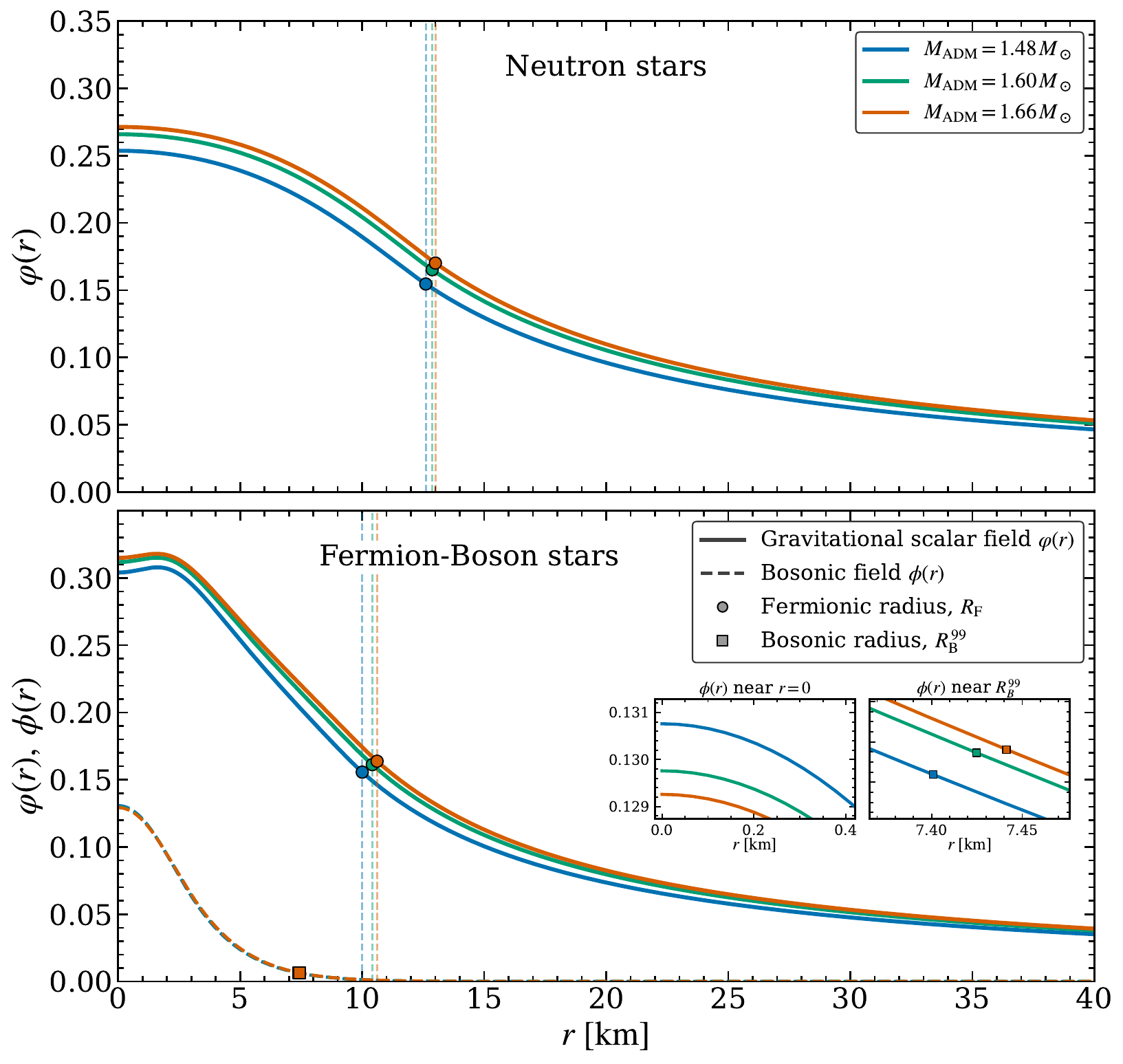}
    \caption{Spatial profiles of bosonic field $\phi(r)$ and gravitational scalar field $\varphi(r)$, for three pairs of scalarized configurations. The upper panel shows three pure neutron stars, while the lower panel displays their energetically favored scalarized fermion-boson star counterparts. Each pair has the same total gravitational mass, indicated by colors in a legend box. All solutions are constructed with $\beta=-7$. Circles (squares) mark the fermionic (bosonic) radius $\RF$ ($\RB$). Vertical lines show the corresponding fermionic surfaces. The inset panels magnify bosonic profiles near the origin and around $\RB$. The bosonic core of fermion-boson stars tends to suppress $\varphi(r)$ relative to pure neutron stars.
    }
    \label{fig:varphis_beta-7}
\end{figure}

\textit{Bosonic Core Screening.—}
In all favored FBS configurations within the observational window, the bosonic component forms a compact \textit{bosonic core} embedded inside the fermionic fluid; see Fig.~\ref{fig:3D_plot}. This central mass concentration redistributes the source trace $\tilde{T}$ of the scalar field equation $\tilde{\Box}\varphi = -4\pi G \alpha(\varphi)\tilde{T}$. In Fig.~\ref{fig:varphis_beta-7} we compare gravitational scalar field profiles, $\varphi(r)$, corresponding to pure scalarized NSs and energetically favored FBSs (featuring bosonic cores) of identical total mass across coupling values $\beta \in \{-6, -7, -8\}$.  The gravitational scalar field $\varphi(r)$ in FBSs exhibits a steeper spatial decay at large distances compared to pure scalarized NSs of the same total mass. This steeper falloff tends to suppress the scalar charge $Q_\varphi$, effectively acting as a \textit{bosonic screening mechanism} that should damp dipolar scalar radiation and alleviate constraints from binary pulsars \cite{Freire:2012mg,Antoniadis:2013pzd,Shao:2017gwu,Tuna:2022qqr}. In the representative case of $\beta=-7$, the scalar charge difference can be up to $0.3685M_\odot$. This behavior is not specific to the case $\beta=-7$, as shown in the End Matter (Fig.~\ref{fig:varphis_other_betas} and Table~\ref{tab:equal_adm_three_masses_beta6_beta7_beta8}).

Current bounds on massless DEF theory~\cite{Bertotti:2003rm,Freire:2012mg,Antoniadis:2013pzd,Shao:2017gwu,Niu:2021nic,Tuna:2022qqr} can be relaxed by introducing a mass for the DEF scalar, which suppresses its exterior field~\cite{PhysRevD.92.124016,Ramazanoglu:2016kul,Yazadjiev:2016pcb,Doneva:2016xmf,Sperhake:2017itk,Staykov:2018hhc,Rosca-Mead:2019seq,Rosca-Mead:2020bzt,Huang:2021tpu,Kuan:2023hrh,Asakawa:2023obq,Degollado:2024oyo,Lam:2024wpq}. Although the gravitational scalar remains massless in our model, the energetically favored configurations in our observational window exhibit a qualitatively similar screening effect. This reduction stems from the redistribution of the source trace $\tilde T$ rather than a Yukawa-type exponential suppression. This does not by itself evade binary-pulsar constraints, since the associated dipolar radiation must still be evaluated explicitly. Nevertheless, it suggests that mixed stars may support appreciable internal scalarization alongside a comparatively weaker exterior scalar signature. Establishing whether this mechanism can genuinely reconcile scalarization with current bounds will require calculations of scalar charges and binary dynamics.

\textit{Testing Prospects.—}
Our analysis reveals a regime of potential tension for massless DEF gravity. In the upper-mass region of the observational window ($M_{\text{ADM}} \gtrsim 1.8M_\odot$), pure scalarized NS solutions do not exist; see Fig.~\ref{fig:window-several-betas}. In this domain, massless DEF gravity exclusively predicts scalarized FBS configurations as energetically favored states. If future multi-messenger observations definitively establish that compact objects in this mass range are unscalarized \emph{pure} NSs (for instance employing next-generation observatories such as the Einstein Telescope \cite{Punturo:2010zza,Hild:2010id} and Cosmic Explorer \cite{Reitze:2019iox,LIGOScientific:2016wof}), this prediction could be leveraged to place unprecedented constraints on—or even rule out—the coexistence of massless DEF gravity and ultralight dark matter condensates.

\textit{Conclusions.—}
We have explored the observational viability of scalarized FBSs in massless DEF scalar-tensor gravity, utilizing the realistic MPA1 nuclear equation of state and a self-interacting complex scalar field as a dark matter model. 
Our main findings are threefold:
\begin{enumerate}
    \item \textbf{Catalysis of scalarization.} Bosonic dark matter lowers the critical coupling threshold $|\beta_{\text{crit}}|$, facilitating spontaneous scalarization across the range of the observational window; see Fig.~\ref{fig:beta_crit}.
    \item \textbf{Energetically favored models and viability.} Energetically favored models inside the observational window are \emph{scalarized} FBSs that feature bosonic dark matter cores. The compactifying effect of bosonic cores restores the viability of strong coupling regimes that would otherwise be ruled out by NS radius bounds; see Fig.~\ref{fig:window-several-betas}.
    \item \textbf{Screening and testing prospects.} Bosonic dark matter cores tend to suppress the scalar charge $Q_\varphi$ relative to pure scalarized NSs; see Fig.~\ref{fig:varphis_beta-7}. This effect partially mimics that of a massive gravitational scalar field in DEF theory. At the same time, the prediction of scalarized FBSs in regions lacking pure scalarized NSs offers an observational target to test this scenario.
\end{enumerate}

Our results suggest that bosonic dark matter cannot be ignored when testing modified gravity in the strong-field regime. We have obtained similar results for three realistic equations of state, and the Companion Paper~\cite{CMjOnFs2026b} finds the same trends for different boson masses. Mapping the dependence on the self-interaction parameter $\lambda$, extending the analysis to broader $\mu$ ranges, and assessing the stability and formation of these configurations are natural directions for future work.\\ \\ \\

\begin{acknowledgments}
JCM is funded by the UNAM-DGAPA Postdoctoral Fellowship Program, and by the UNAM-PAPIIT project No. IN108526. SF is supported by the Generalitat Valenciana and the European Social Fund through the APOSTD 2025 fellowship (CIAPOS/2024/461), and partially by MICIU/AEI and FEDER/EU (grant PID2024-159689NB-C21) and the EU Horizon Europe programme (NewFunFiCO, grant No. 101086251). Numerical simulations were carried out at the UNAM-\textsc{Lamod} facility.
\end{acknowledgments}

\bibliographystyle{apsrev4-2}
\bibliography{biblio}

\appendix
\section{End Matter}
\begin{figure}[h!]
    \centering
     \includegraphics[width=\columnwidth]{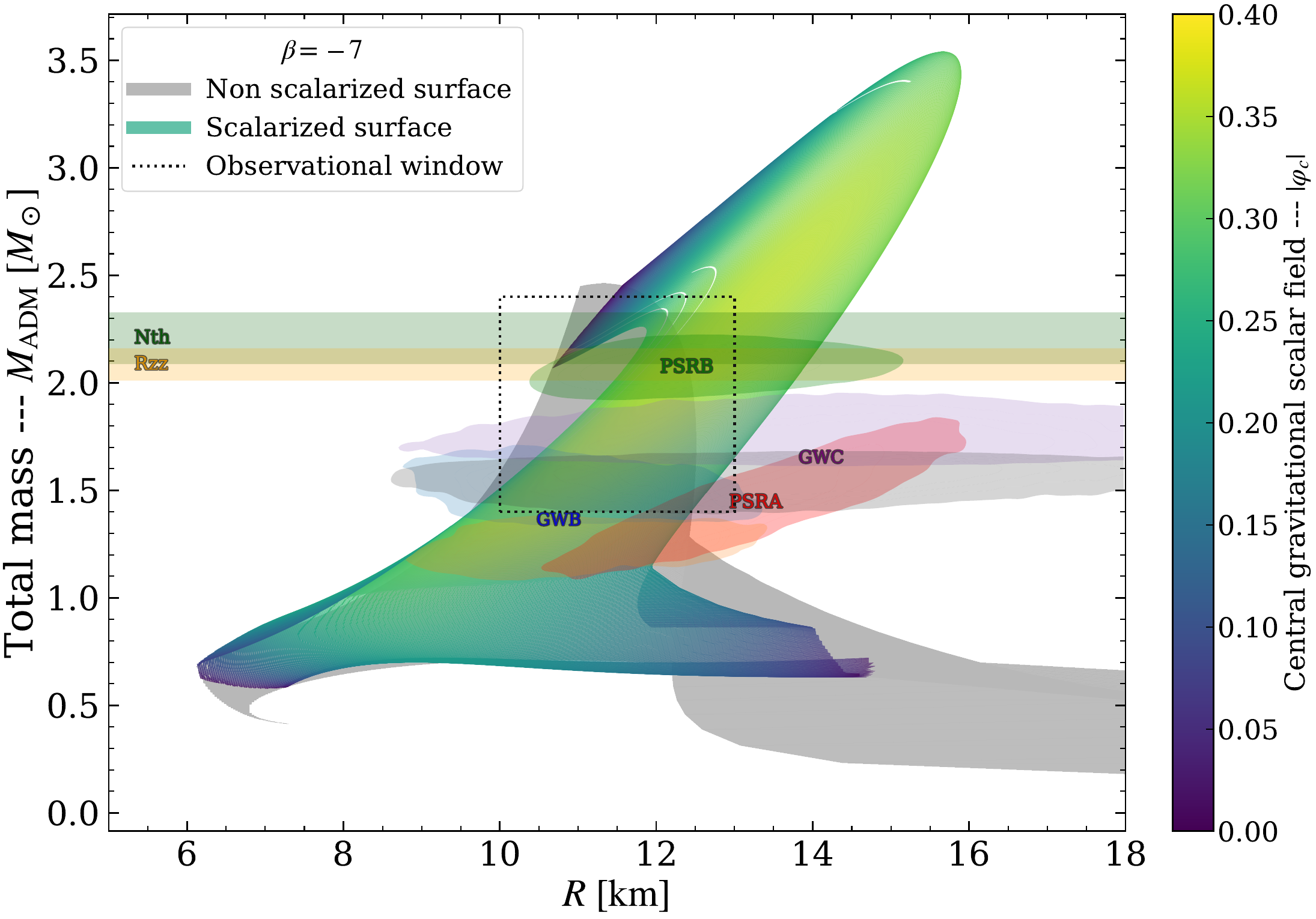}
    \caption{Mass-radius diagram for FBSs constructed with the MPA1 equation of state, $\beta=-7$, $\mu^2=1$, and $\Lambda_4=50$, shown together with representative observational constraints. The gray background represents the non-scalarized solutions, while the colored surface corresponds to scalarized solutions; its color scale denotes the absolute value of the central gravitational scalar field, $|\varphi_c|$. The shaded posterior regions labeled PSRA and PSRB correspond to the NICER mass-radius constraints for PSR J0030+0451 \cite{miller2019psr} and PSR J0740+6620 \cite{Legred:2021hdx}, respectively, whereas GWB and GWC denote constraints inferred from GW170817 and GW190425 \cite{abbott2017gw170817,LIGOScientific:2020aai}. Horizontal bands labeled Rzz and Nth  indicate the corresponding maximum-mass constraints given in \cite{Rezzolla:2017aly,Nathanail:2021tay}. Dotted lines mark the window $10\leq R/{\rm km}\leq13$ and $1.4\leq M_{\rm ADM}/M_\odot\leq2.4$ used in our parameter-space analysis.}
    \label{fig:observational_bounds}
\end{figure}

\begin{figure*}[htbp]
    \centering
    \includegraphics[width=0.49\textwidth]{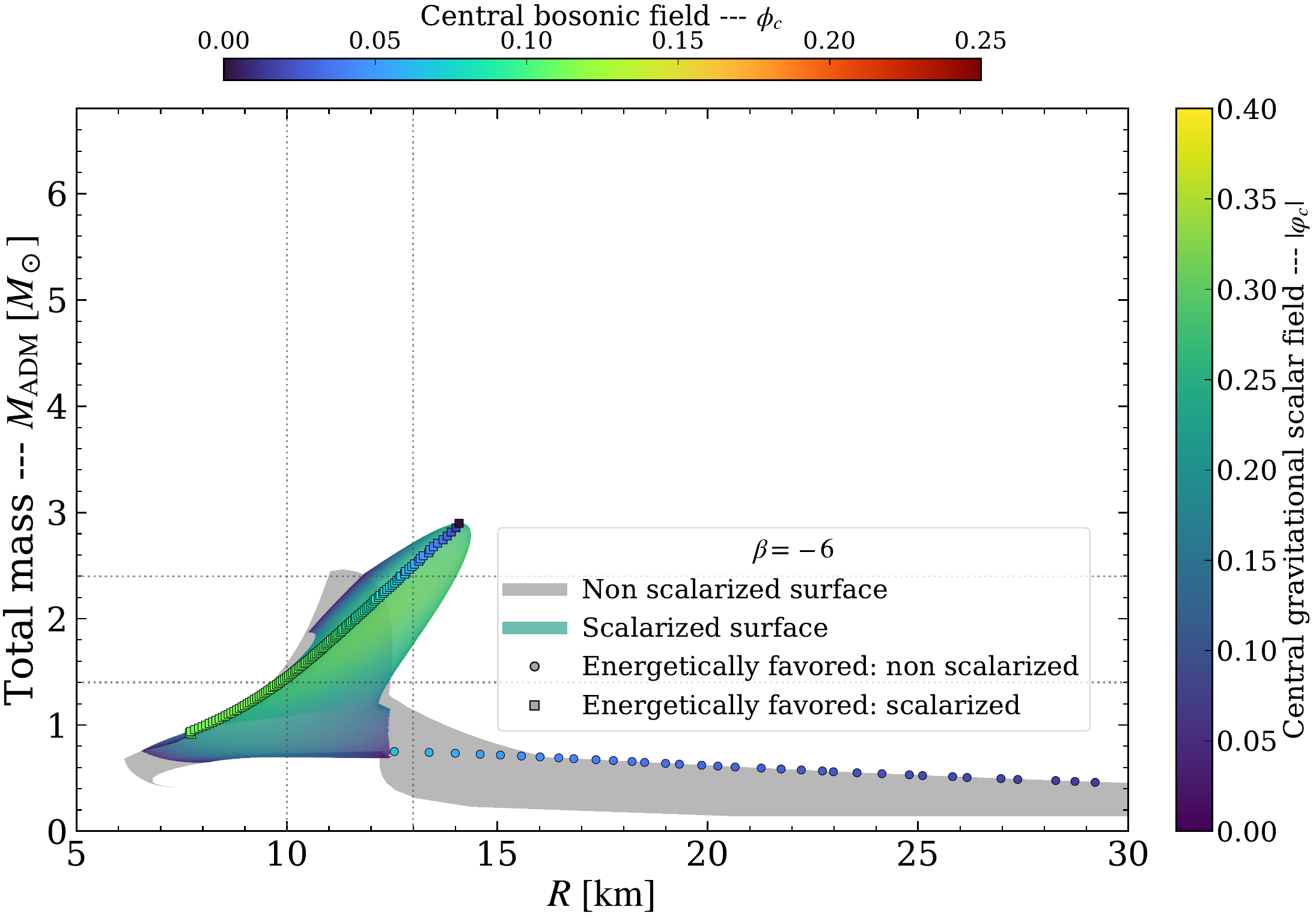}
    \hfill
    \includegraphics[width=0.49\textwidth]{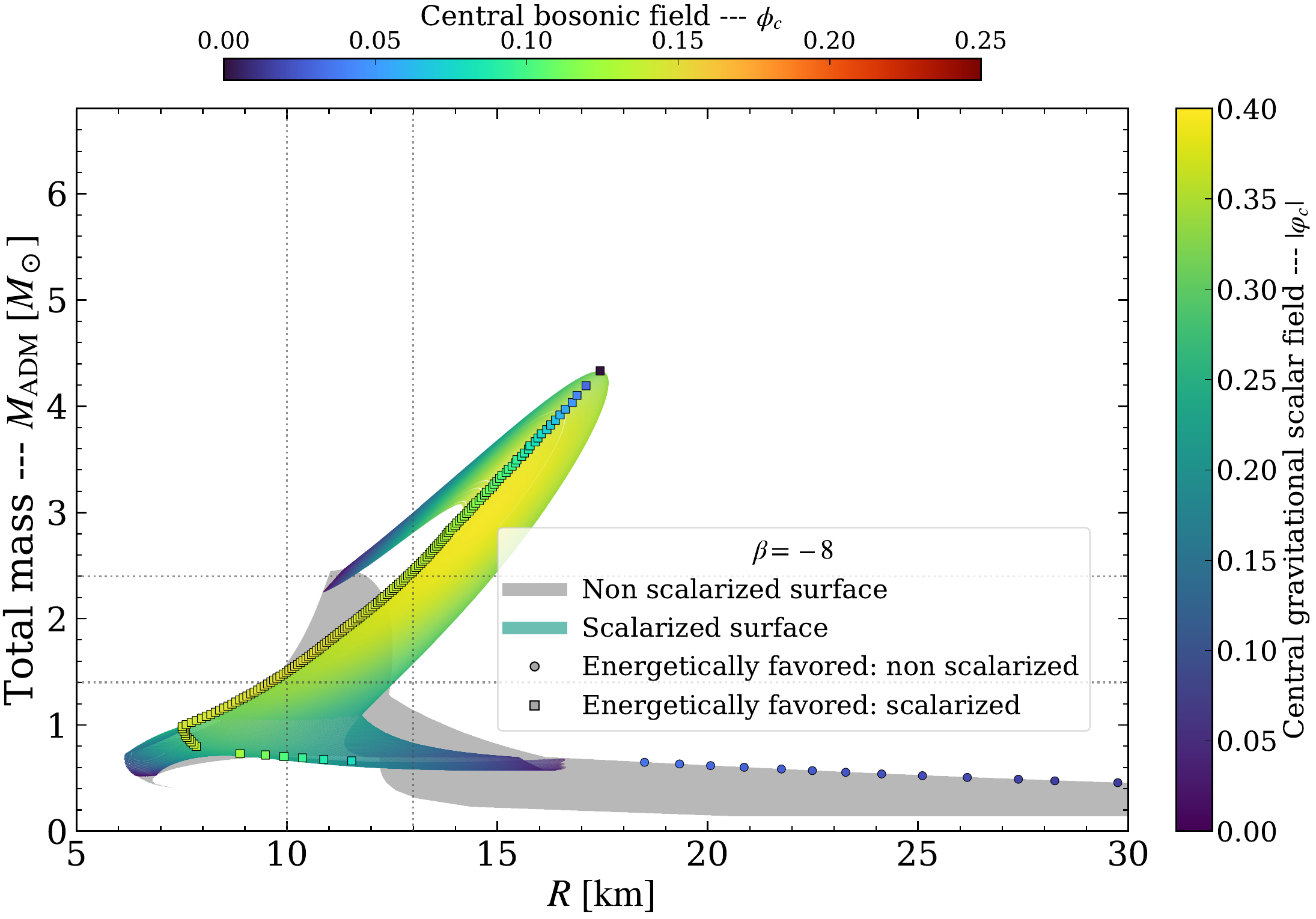}

    \vspace{0.4em}

    \includegraphics[width=0.49\textwidth]{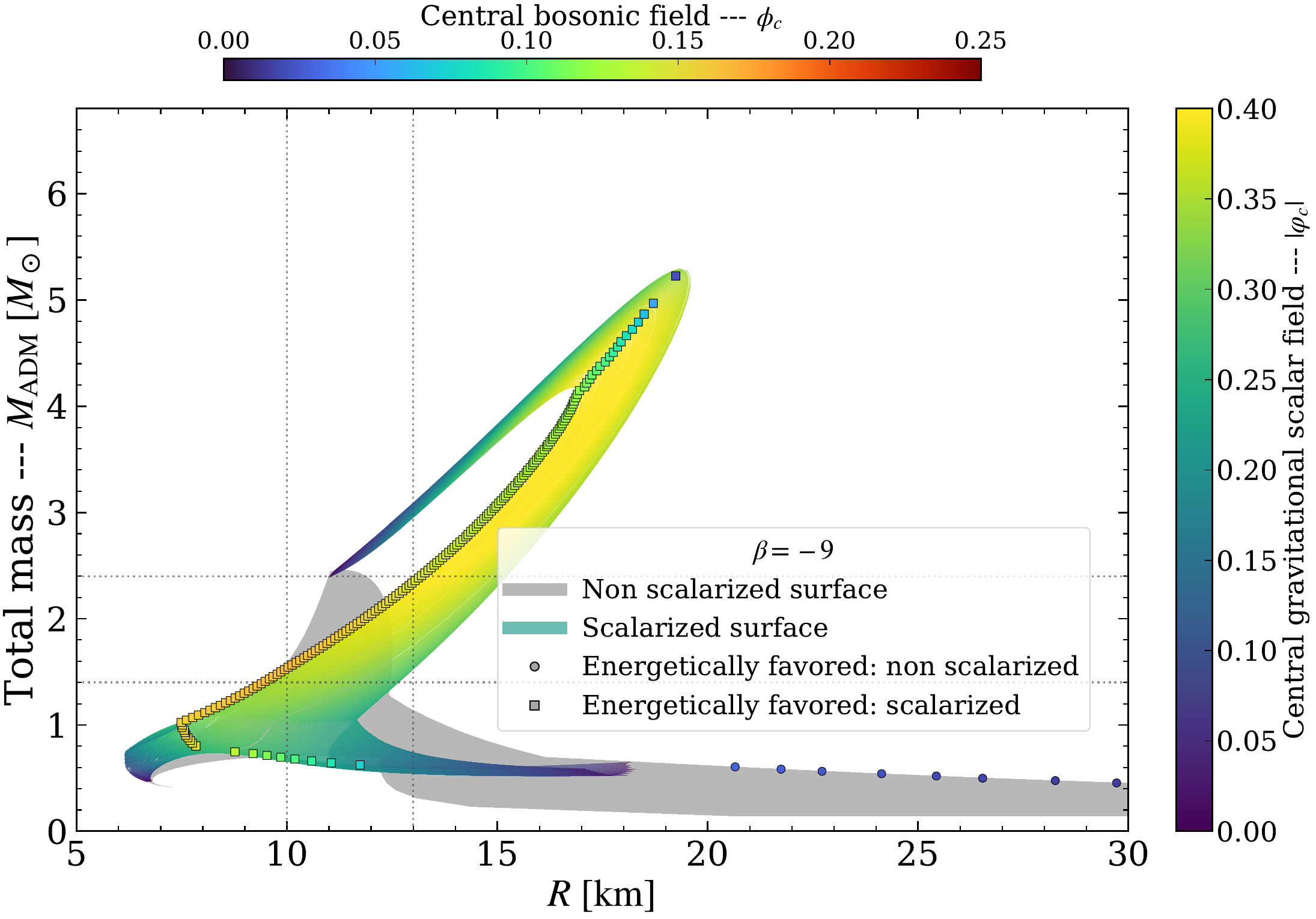}
    \hfill
    \includegraphics[width=0.49\textwidth]{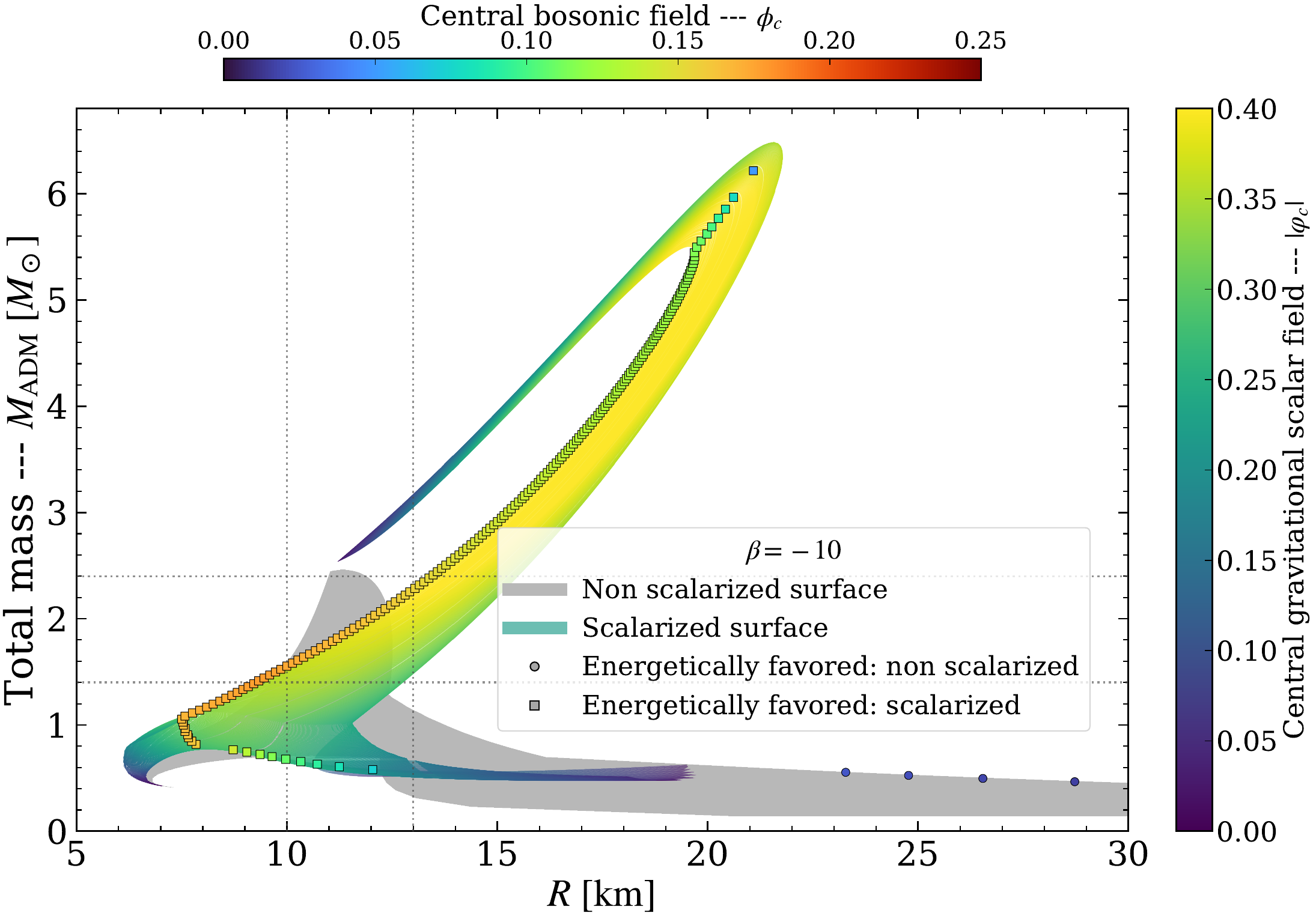}

    \caption{
   Mass-radius diagrams for FBSs with $\beta\in\{-6,-8,-9,-10\}$, from top left to bottom right. The case of $\beta=-7$ is shown in Fig.~\ref{fig:M-R-beta-7}. All configurations are constructed with the MPA1 equation of state, $m_{\rm B}=1.34\times10^{-10}$ eV, and $\Lambda_4=50$. Gray and colored surfaces represent non-scalarized and scalarized solutions, respectively, with the latter colored according to $|\varphi_c|$. Dots and squares denote, respectively, energetically favored non-scalarized and scalarized configurations; their color indicates $\phi_c$. Straight dotted lines delimit the observational window. The same mass range, $0\leq M_{\rm ADM}/M_\odot\leq6.8$, is used in every panel to appreciate how the maximum mass allowed by scalarized solutions increases as $\beta$ becomes more negative.}
    \label{fig:M-R-different_beta}
\end{figure*}

\textbf{Critical curve for the onset of scalarization.}
Regarding the construction of Fig.~\ref{fig:beta_crit}, the parameters $\rho_{0c}$, $\phi_c$, and $\beta$ jointly determine whether a configuration lies within the observationally relevant mass-radius region shown in Fig.~\ref{fig:observational_bounds}, which is independent of the theory parameters. For each value of $\beta\in\{-6,-7,-8,-9,-10\}$, we identify the ranges of $\rho_{0c}$ and $\phi_c$ that produce configurations inside the prescribed mass-radius window. We then take the envelope—or union—of these parameter intervals over all the values of $\beta$ under consideration. This procedure yields the maximal domain associated with observationally relevant configurations, $0.758\times10^{15}\,{\rm g\,cm^{-3}}\lesssim\rho_{0c}\lesssim 1.690\times10^{15}\,{\rm g\,cm^{-3}}$, and $0\leq\phi_c\lesssim 0.1733$.
The two limiting values of $\rho_{0c}$ are then used as lower and upper density boundaries in Fig.~\ref{fig:beta_crit}. Finally, the intersection boundary defines the critical curve $\beta_{\rm crit}(\phi_c)$ within the observational window.

\textbf{Energetically favored configurations.}
Given the total rest mass of fermions and bosons, $M_{\rm 0,F}$ and $M_{\rm 0,B}$, respectively, contained in a FBS, we define the total rest mass, $M_0 = M_{\rm 0,F} + M_{\rm 0,B}$; see details in the Companion Paper~\cite{CMjOnFs2026b}. For a given value of $M_0$, we find the configuration that minimizes the total gravitational mass,
\begin{equation}
    M_{\rm ADM}^{\rm min}(M_0)
    =
    \min_{\{\,{\rm models}\,|\,M_0={\rm const.}\,\}}
    M_{\rm ADM}.
\end{equation}
Figures~\ref{fig:M-R-beta-7} and~\ref{fig:M-R-different_beta} show a series of mass-radius relations for coupling parameters $\beta\in\{-6,-7,-8,-9,-10\}$. Scalarized solutions are shown with a colormap encoding $|\varphi_c|$, while non-scalarized solutions appear in gray. Dots and squares correspond to configurations that minimize $M_{\rm ADM}$ for a fixed $M_0$, and their colormap encodes $\phi_c$. Dots (squares) indicate minima when they correspond to non-scalarized (scalarized) solutions.
The set of energetically preferred, scalarized solutions traces a sequence from the low- to high-mass regime. At the high-mass end, the preferred configurations are associated with relatively small values of $\phi_c$; moving toward intermediate masses, the minimizing solutions shift toward larger bosonic amplitudes, before $\phi_c$ decreases again along the lower-mass part of the sequence. These energetically preferred configurations do not simply lie within the scalarized domain, but predominantly trace its strongly scalarized region. This indicates that, over a substantial portion of the mass-radius diagram, highly scalarized FBSs are energetically favored at fixed $M_0$.

\begin{figure*}[htbp]
	\centering
    \includegraphics[width=0.45\textwidth]{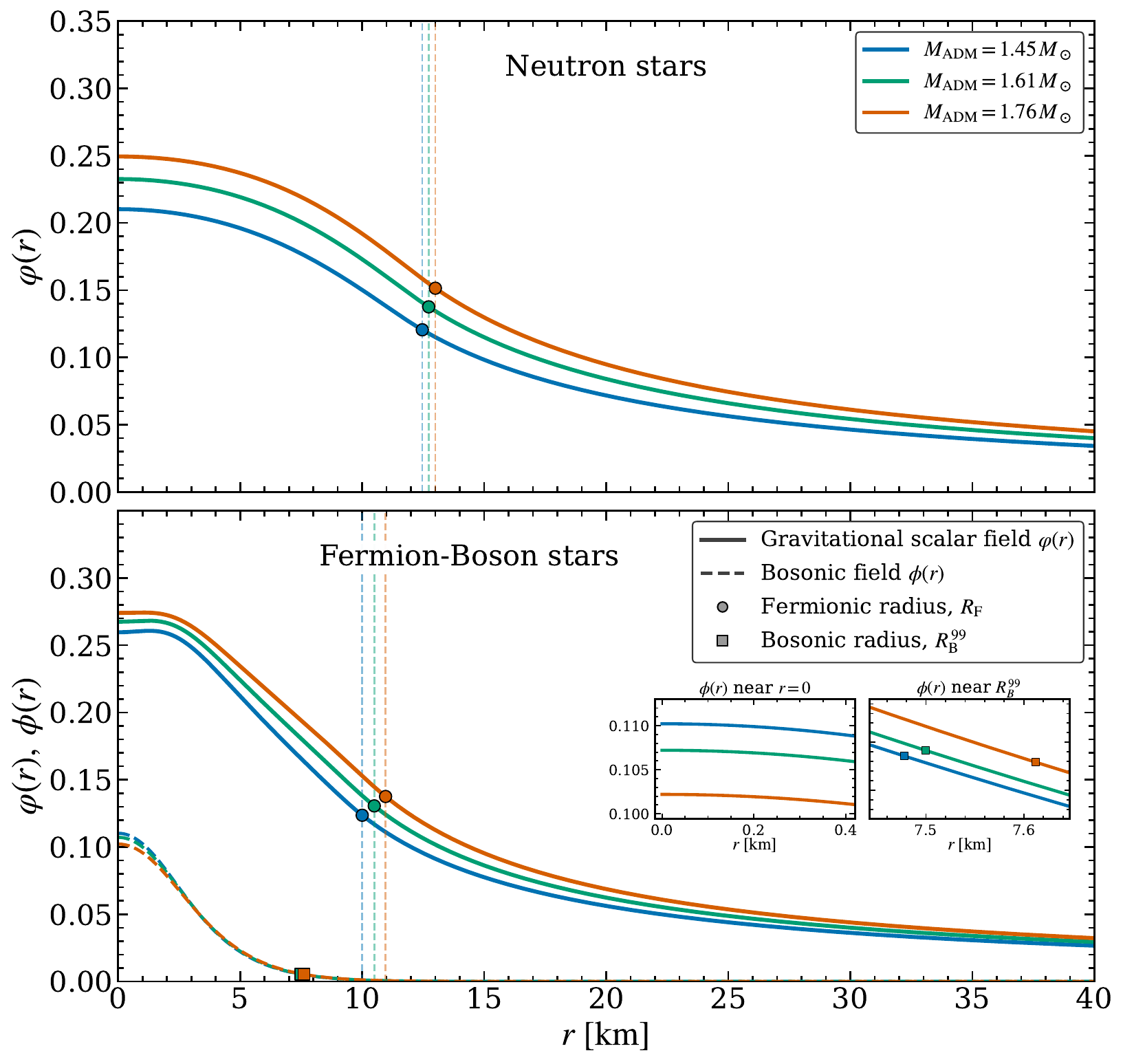}
    	\vspace{0.2cm}
    \includegraphics[width=0.45\textwidth]{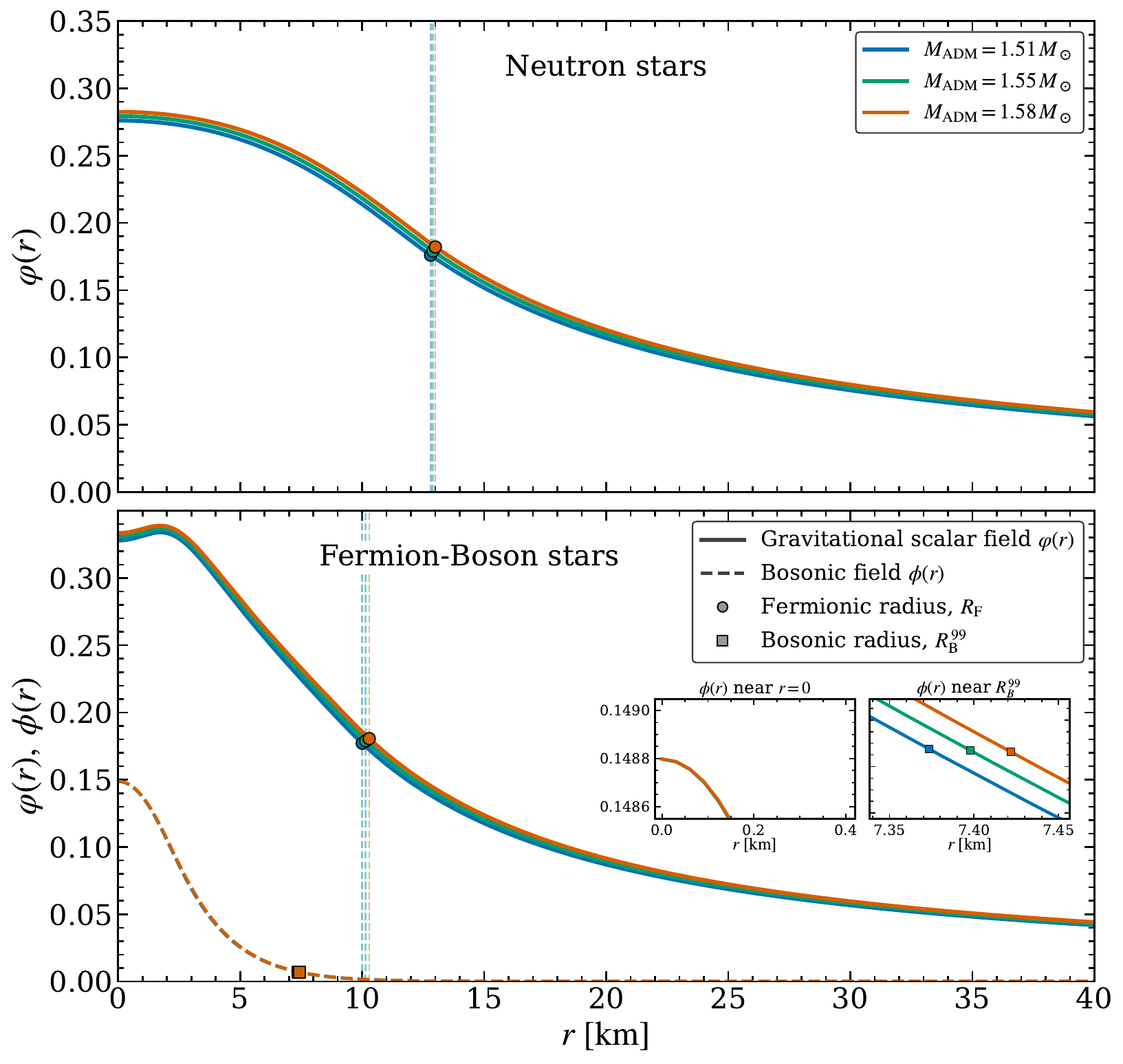}

    \caption{Same as Fig.~\ref{fig:varphis_beta-7}, for $\beta=-6$ (left) and $\beta=-8$ (right), showing that gravitational scalar field suppression by FBSs with bosonic cores persists for different choices of coupling parameter $\beta$. }
    \label{fig:varphis_other_betas}
\end{figure*}

\textbf{Gravitational scalar field suppression.}
From Fig.~\ref {fig:varphis_beta-7} we infer that FBSs with bosonic cores can contribute to screen DEF gravity effects for $\beta=-7$. Figure~\ref{fig:varphis_other_betas} shows that the screening persists for $\beta\in\{-6,-8\}$. Correspondingly, in Table~\ref{tab:equal_adm_three_masses_beta6_beta7_beta8} we provide the parameters corresponding to the cases of $\beta\in\{-6,-7,-8\}$.

\begin{table*}[h!]
\centering
\caption{
Parameters of scalarized pure NSs and their energetically favored scalarized FBS (E-F FBS) counterparts, for coupling $\beta\in\{-6,-7,-8\}$. Within each pair, both configurations have the same total mass $M_{\rm ADM}$. Although masses are displayed to two decimal places, the numerically computed masses within each pair
agree to at least $5\times10^{-8}\,M_\odot$. All configurations use the MPA1 equation of state, $m_{\rm B}=1.34\times10^{-10}$ eV, and $\Lambda_4=50$. $\rho_{0c}$ denotes central rest mass density; $p_c$ denotes central pressure; $\phi_c$ is central bosonic field; $\varphi_c$ is central gravitational scalar field; $\omega/\mu$ is normalized bosonic field frequency; $\RF$ ($\RB$) is fermionic (bosonic) radius; and $Q_\varphi$ denotes scalar charge. Units are indicated in column headers.
}
\label{tab:equal_adm_three_masses_beta6_beta7_beta8}
\renewcommand{\arraystretch}{1.05}
\setlength{\tabcolsep}{2.5pt}
\resizebox{0.9\textwidth}{!}{%
\begin{tabular}{c|cc c c c c c c c c}
\hline\hline
$\beta$ & $M_{\rm ADM}$ & Model & $\rho_{0c}$ & $p_c$ &
$\phi_c$ & $\varphi_c$ & $\omega/\mu$ &
$\RF$ & $\RB$ & $|Q_\varphi|$ \\
& $[M_\odot]$ & &
$[10^{15}\,{\rm g\,cm^{-3}}]$ &
$[10^{-4}\,M_\odot\,{\rm km}^{-3}]$ &
& & & $[{\rm km}]$ & $[{\rm km}]$ & $[M_\odot]$ \\
\hline
 & 1.45 & Pure NS & 0.770 & 0.802 & 0.0000 & 0.2102 & --- & 12.46 & 0.00 & 0.8942 \\
 & 1.45 & E-F FBS & 1.399 & 5.301 & 0.1102 & 0.2596 & 0.5512 & 10.00 & 7.48 & 0.6957 \\
\cline{2-11}
\raisebox{-0.55\baselineskip}{$ -6 $} & 1.61 & Pure NS & 0.807 & 0.955 & 0.0000 & 0.2326 & --- & 12.72 & 0.00 & 1.0418 \\
 & 1.61 & E-F FBS & 1.419 & 5.513 & 0.1072 & 0.2675 & 0.5233 & 10.49 & 7.50 & 0.7647 \\
\cline{2-11}
 & 1.76 & Pure NS & 0.842 & 1.097 & 0.0000 & 0.2494 & --- & 13.00 & 0.00 & 1.1703 \\
 & 1.76 & E-F FBS & 1.418 & 5.507 & 0.1022 & 0.2740 & 0.5003 & 10.96 & 7.61 & 0.8348 \\
\hline
 & 1.48 & Pure NS & 0.812 & 0.971 & 0.0000 & 0.2537 & --- & 12.61 & 0.00 & 1.2189 \\
 & 1.48 & E-F FBS & 1.515 & 6.555 & 0.1308 & 0.3041 & 0.5228 & 10.00 & 7.40 & 0.9190 \\
\cline{2-11}
\raisebox{-0.55\baselineskip}{$ -7 $} & 1.60 & Pure NS & 0.836 & 1.070 & 0.0000 & 0.2660 & --- & 12.87 & 0.00 & 1.3350 \\
 & 1.60 & E-F FBS & 1.537 & 6.810 & 0.1298 & 0.3118 & 0.5028 & 10.42 & 7.42 & 0.9899 \\
\cline{2-11}
 & 1.66 & Pure NS & 0.846 & 1.117 & 0.0000 & 0.2714 & --- & 13.00 & 0.00 & 1.3915 \\
 & 1.66 & E-F FBS & 1.549 & 6.942 & 0.1293 & 0.3149 & 0.4932 & 10.61 & 7.44 & 1.0230 \\
\hline
 & 1.51 & Pure NS & 0.843 & 1.105 & 0.0000 & 0.2762 & --- & 12.80 & 0.00 & 1.4814 \\
 & 1.51 & E-F FBS & 1.607 & 7.616 & 0.1488 & 0.3279 & 0.4987 & 10.00 & 7.37 & 1.0982 \\
\cline{2-11}
\raisebox{-0.55\baselineskip}{$ -8 $} & 1.55 & Pure NS & 0.849 & 1.132 & 0.0000 & 0.2794 & --- & 12.90 & 0.00 & 1.5218 \\
 & 1.55 & E-F FBS & 1.617 & 7.731 & 0.1488 & 0.3307 & 0.4925 & 10.15 & 7.40 & 1.1255 \\
\cline{2-11}
 & 1.58 & Pure NS & 0.855 & 1.160 & 0.0000 & 0.2826 & --- & 13.00 & 0.00 & 1.5628 \\
 & 1.58 & E-F FBS & 1.627 & 7.849 & 0.1488 & 0.3334 & 0.4864 & 10.29 & 7.42 & 1.1529 \\
\hline\hline
\end{tabular}%
}
\end{table*}

\end{document}